\documentclass[aps,twocolumn,pra,superscriptaddress,amsmath,showpacs,tightenlines,pdflatex,longbibliography]{revtex4-1}
\usepackage{amssymb}
\usepackage{amsmath}
\usepackage{dcolumn}
\usepackage{graphicx}
\usepackage{mathrsfs}
\usepackage{appendix}
\usepackage{graphicx}
\usepackage{booktabs}
\usepackage{color}

\usepackage{url}
\usepackage[colorlinks]{hyperref}
\hypersetup{%
    plainpages=true,
    breaklinks=true,
    hypertexnames=false,
    pageanchor=true,
    colorlinks=true,
    linkcolor={blue},
    citecolor={blue},
    urlcolor={blue},
    anchorcolor={black}
}

\begin{document}

\title{Generation and stabilization of entangled coherent states for the
vibrational modes of a trapped ion}
\author{Zhi-Rong Zhong}
\author{Xin-Jie Huang}
\author{Zhen-Biao Yang}
\author{Li-Tuo Shen}
\author{Shi-Biao Zheng}
\email{t96034@fzu.edu.cn}
\address{Fujian Key Laboratory of Quantum Information and Quantum Optics, College of\\
Physics and Information Engineering, Fuzhou University, Fuzhou, Fujian\\
350116, China}
\date{\today }

\begin{abstract}

We propose a scheme for preparation of entangled coherent states for the
motion of an ion in a two-dimensional anisotropic trap. In the scheme, the
ion is driven by four laser beams along different directions in the ion trap
plane, resulting in carrier excitation and couplings between the internal
and external degrees of freedom. When the total quantum number of the
vibrational modes initially has a definite parity, the competition between
the unitary dynamics and spontaneous emission will force the system to
evolve to a steady state, where the vibrational modes are in a two-mode cat
state. We show that the method can be extended to realization of entangled
coherent states for three vibrational modes of an ion in three-dimensional
anisotropic trap.

\end{abstract}

\maketitle

\section{INTRODUCTION}

The superposition principle, distinguishing between the quantum and
classical worlds, lies at the heart of quantum mechanics. Superimposing two
state components gives arise to new nonclassical effects due to the quantum
interference between these components. This is well exemplified by
superpositions of macroscopically distinguishable states, known as
Schrodinger cat states [1]. In quantum optics, cat states are defined as
superpositions of two distinct quantum states closest to classical ones$-$%
coherent states with different phases or amplitudes. Though formed by
quasiclassical states, cat states can exhibit stongly nonclassical
behaviors, i.e, negativity of the quasiprobability distribution in phase
space [2]. These superposition states are central to exploration of the
fuzzy quantum-classical boundary, and useful for redundant encoding required
for quantum error correction [3-6]. Such states have been experimentally
generated in various systems, including cavity quantum electrodynamics (QED)
[2,7], superconducting circuits [8-13], and ion traps [14-16].

The application of the superposition principle to composite systems will
give arise to a more striking quantum phenomenon, i.e., entanglement,
originally introduced by Einstein et al. to question the completeness of
quantum mechanics. Among various entangled states, entangled coherent states
of two harmonic oscillators [17], also referred to as two-mode cat states
[18], are strikingly interesting. The entanglement between quasiclassical
state components leads to interesting nonclassical properties, such as
two-mode squeezing, violations of Cauchy-Schwartz inequality, quantum
interference features in four-dimensional phase space, and violations of
Bell inequalities. Apart from fundamental interest, these entangled states
have practical applications in quantum information processing [19] and
quantum metrology [20]. Schemes have been proposed for producing such states
for two mesoscopic fields through reservoir engineering in cavity QED
[21,22] and circuit QED [23]. Recently, these states have been produced in a
circuit [24], where the microwave fields stored in two three-dimensional
cavities were entangled by dispersively coupling them to a superconducting
qubit.

Ion traps represent another qualified candidate for quantum state
engineering and quantum information processing. The prominent properties of
this system is that the damping of the vibrational modes is extremely weak,
and these bosonic modes can be controllably coupled to the electronic state
via laser driving. In addition to the Schrodinger cat states [14-16],
squeezed states [25], Fock states [25], and superpositions of Fock states
[26] have been experimentally produced for one vibrational mode of a trapped
ion. Entanglement consisting of zero- and one-phonon states has also been
observed for two vibtational modes, each belonging to one pair of trapped
ions [27]. On the other hand, proposals have been suggested for generating
various entangled states for two vibrational degrees of freedom of a trapped
ion, including pair coherent states [28], superpositions of pair coherent
states [29], SU(1,1) intelligent states [30], two-mode squeezed pair
coherent states [31], entangled coherent states [32,33], and arbitrary
superpositions of two-mode Fock states [34-38]. The schemes proposed in
Refs. [32] and [33] are relying on measurement of the ionic internal state
to conditionally project the vibrations to the desired state after suitable
laser driving, while those of Refs. [34], [35], [36], [37], and [38] allow
for deterministic generation of arbitraty entangled states in the Fock state
basis through step-by-step procedures, following which any entangled
coherent state can be approximately realized by approaching its Fock state
expansion, with the number of required operations increasing with the
average quantum number of each mode. We here propose an unconditional scheme
for generating such states of the two vibrational modes of an ion confined
in a two-dimensional anisotropic harmonic trap. In our scheme, the ion is
driven by four lasers of different frequencies on the xy plane. With
suitable setting of the parameters and directions of these lasers, the
system steady state is given by the product of the electronic ground state
with a vibrational two-mode cat state if the two vibrational modes initially
have a definite joint parity. Unlike the previous schemes [32-38], our
scheme requires neither the projective measurement of the ionic electronic
state nor a sequence of operations achieved by frequently tuning the
parameters of the driving laser. Another distinct feature of our scheme is
that it not only allows for the generation of entangled coherent states, but
also can be used to stabilize these states.

The paper is organized as follows. In Sec. 2, we present the theoretical
model, in which the internal state of a trapped ion is coupled to its
vibrational modes along two mutually perpendicular directions by laser
driving. The unitary dynamics associated with this coupling, together with
the decaying of the internal state, forces the system to evolve to the
steady state, where the two vibrational modes are in a two-mode cat state.
In Sec 3, we present numerical simulations of the fidelities of the
vibrational states to the desired cat states with the dissipation of the
vibrational modes being included, confirming the validity of our scheme. We
also display the Wigner functions of the vibrational modes in the steady
state, which show good agreement with the ideal cat states. In Sec. 4, we
show that the scheme can be generalized to drive three vibrational modes of
an ion confined in an anisotropic three-dimensional harmonic trap into a
three-mode cat state. A summary appears in Sec. 5.

\section{THEORETICAL MODEL}

We consider an ion trapped in an anisotropic harmonic potential, with the
vibrational frequencies along the x and y axes are $\nu _{x}$ and $\nu _{y}$%
, respectively. The transition between the electronic ground state $%
\left\vert g\right\rangle $ and one excited state $\left\vert e\right\rangle
$ are driven by four laser beams of frequencies $\omega _{0}-2\omega _{x}$, $%
\omega _{0}-2\omega _{y}$, $\omega _{0}-\omega _{x}-\omega _{y}$, and $%
\omega _{0}$, where $\omega _{0}$ is the transition frequency between $%
\left\vert g\right\rangle $ and $\left\vert e\right\rangle $. The first two
are aligned along the x and y axes, while the other two are at angles of $%
\pi /4$ and $-\pi /4$ to x axis, respectively. In the rotating-wave
approximation, the Hamiltonian for this system is given by (setting $\hbar
=1 $)
\begin{equation}
H=\omega _{x}a^{\dagger }a+\omega _{y}b^{\dagger }b+\omega
_{0}S_{z}+[\lambda E^{+}(x,y,t)S^{+}+H.c.],
\end{equation}%
where $a^{\dagger }$ ($b^{\dagger }$) and $a$ ($b$) are the creation and
annihilation operators for the vibrational modes along the x and y axes and
of frequencies $\omega _{x}$ and $\omega _{y}$, respectively, $S^{+},$ $%
S^{-},$ and $S_{z}$ are the raising, lowering, and inversion operators for
the electronic diplole transition, and $\lambda $ is the transition
frequency and dipole matrix element. $E^{+}(x,t)$ is the positive part of
the classical driving fields

\begin{eqnarray}
E^{+}(x,t) =&& E_1e^{-i[(\omega _0-2\omega _x)t-k_1x +\phi
_1]} \nonumber\\&& +E_2e^{-i[(\omega _0-2\omega _y)t-k_2y+\phi _2]}  \nonumber\\&& \label{2} +E_3e^{-i[(\omega _0-\omega _x-\omega _y)t-k_3(x+y)/\sqrt{2}+\phi_3]} \nonumber\\&& + E_0e^{-i[\omega _0t-k_0(x-y)/\sqrt{2}+\phi _0]},
\end{eqnarray}
where $E_n,$ $\phi _n,$ and $k_n$ ($n=0,1,2,3$) are the amplitudes, phases,
and wave vetctors of the $n$th driving field, respectively. The position
operators $x$ and $y$ can be expressed by $x=\sqrt{1/(2\omega _xM)}%
(a+a^{\dagger })$ and $y=\sqrt{1/(2\omega _yM)}(b+b^{\dagger })$, with $M$
being the mass of the trapped ion.

In the resolved sideband limit the vibrational frequencies $\omega _{x}$ and
$\omega _{y}$ are much larger than other characteristic frequencies of the
problem. Then the interactions of the ion with lasers can be treated using
nonlinear Jaynes-Cummings model [39]. In this case the Hamiltonian for such
a system , in the interaction picture, is given by

\begin{widetext}
\begin{align}
&&H_i = \sum_{j=0}^{\infty}\left\{ e^{-\eta _x^2/2}\frac{
(i\eta _x)^{2j+2}}{j!(j+2)!}\Omega _1e^{-i\phi _1}a^{\dagger
j}a^{j+2} \  + \ e^{-\eta _y^2/2}\frac{(i\eta _y)^{2j+2}}{j!(j+2)!}\Omega
_2e^{-i\phi _2}b^{\dagger j}b^{j+2}\right.  \ +
 \sum_{l=0}^{\infty}e^{-(\eta _x^2+\eta_y^2)/4} \cdot  \nonumber\\&& \left[ \frac{(i\eta _x/\sqrt{2})^{2j+1}}{j!(j+1)!}\frac{(i\eta _y/
\sqrt{2})^{2l+1}}{l!(l+1)!}\Omega _3e^{-i\phi _3}a^{\dagger j}a^{j+1}b^{\dagger l}b^{l+1}\right.
\left. \left. +\frac{(i\eta _x/\sqrt{2})^{2j}}{(j!)^2}\frac{(-i\eta _y/\sqrt{2})^{2l}}{(l!)^2}\Omega _0e^{-i\phi _0}a^{\dagger j}a^jb^{\dagger l}b^{l+1}\right] \right\} S^{+} \nonumber\\&& +H.c.,
\end{align}
\end{widetext}
where $\Omega _{n}=\lambda E_{n}$ are the Rabi frequencies of the respective
lasers, and $\eta _{x}=k_{0}/\sqrt{2\omega _{x}M}$ and $\eta _{y}=k_{0}/%
\sqrt{2\omega _{y}M}$ are the Lamb-Dicke parameters associated with the
vibrational modes along the x and y axes, respectively. We here set $%
k_{1}\simeq k_{2}\simeq k_{3}\simeq k_{0}$.

We consider the behavior of the ion in the Lamb-Dicke regime, $\eta
_{x},\eta _{y}\ll 1$. In this limit we can discard the terms with $j>0$ or $%
l>0$. Then the Hamiltonian can be simplified to
\begin{equation}
\begin{array}{c}
H_{i}=\left[ -\frac{\eta _{x}^{2}}{2}e^{-\eta _{x}^{2}/2}\Omega
_{1}e^{-i\phi _{1}}a^{2}-\frac{\eta _{y}^{2}}{2}e^{-\eta _{y}^{2}/2}\Omega
_{2}e^{-i\phi _{2}}b^{2}\right. \\
\left. +e^{-(\eta _{x}^{2}+\eta _{y}^{2})/4}\left( -\frac{\eta _{x}\eta _{y}%
}{2}\Omega _{3}e^{-i\phi _{3}}ab+\Omega _{0}e^{-i\phi _{0}}\right) \right]
S^{+}+H.c.,%
\end{array}%
\end{equation}%
The phase difference $\phi _{n}-\phi _{m}$ is equal to the relative phase
between the $n$th and $m$th driving fields, and the ratio between the Rabi
frequencies $\Omega _{n}$ and $\Omega _{m}$ depends on the ratio between the
amplitudes of these two fields. Therefore, we can choose the relative phases
and amplitudes of the lasers appropriately so that
\begin{eqnarray}
\phi _{1} &=&\phi _{2}=\phi _{3}=\phi _{0},  \nonumber \\
\eta _{x}^{2}e^{-\eta _{x}^{2}/2}\Omega _{1} &=&\eta _{y}^{2}e^{-\eta
_{y}^{2}/2}\Omega _{2}=\frac{\eta _{x}\eta _{y}}{2}e^{-(\eta _{x}^{2}+\eta
_{y}^{2})/4}\Omega _{3}=2\lambda ,\nonumber \\
\end{eqnarray}%
Then we obtain

\begin{equation}
H_{i}=[-\lambda (a+b)^{2}+\varepsilon ]e^{-i\phi _{0}}S^{+}+H.c.
\end{equation}%
where

\begin{equation}
\varepsilon =e^{-(\eta _{x}^{2}+\eta _{y}^{2})/4}\Omega _{0}.
\end{equation}%
The evolution of the vibrational modes is independent of the phase factor $%
e^{-i\phi _{0}}$ in Eq. (6), which can be set to be 1 in the calculation of
the motional state.

The damping of the vibrational modes is so weak that it can be disregarded
and thus the electronic damping is the main decaying process [28-31,40,41].
Then the evolution of the whole system is described by the master equation
for the master operator $\rho $

\begin{equation}
\frac{d\rho }{dt}=-i\left[ H_{i},\text{ }\rho \right] +\frac{\Gamma }{2}%
\left( 2S^{-}\rho S^{+}-S^{+}S^{-}\rho -\rho S^{+}S^{-}\right) ,
\end{equation}%
where $\Gamma $ is the spontaneous decay rate of the excited state of the
ion. In the long time limit, the system will reach the steady state,
satisfying
\begin{equation}
\frac{d\rho _{s}}{dt}=0.
\end{equation}%
In the steady state, the ion will be populated in the ground electronic
state $\left\vert g\right\rangle $ due to the spontaneous emission. As a
consequence, the steady state solution of the master equation (15) can be
rewitten as
\begin{equation}
\rho _{s}=\left\vert g\right\rangle \left\langle g\right\vert \otimes
\left\vert \psi \right\rangle \left\langle \psi \right\vert ,
\end{equation}%
where $\left\vert \psi \right\rangle $ stands for the correlated state of
the motions in the x and y axes. Since the dissipative term has no effect on
the electronic ground state, the condition for the system to reach the
steady state is $\left[ H_{i},\text{ }\rho _{s}\right] =0.$ This leads to
\begin{equation}
\lambda (a+b)^{2}\left\vert \psi \right\rangle =\varepsilon \left\vert \psi
\right\rangle .
\end{equation}%
Any combination of the two-mode coherent states $\left\vert \alpha
_{1}\right\rangle _{a}\left\vert \alpha _{2}\right\rangle _{b}$ and $%
\left\vert -\alpha _{1}\right\rangle _{a}\left\vert -\alpha
_{2}\right\rangle _{b}$ satisfies this equation, where $\left\vert \alpha
_{1}\right\rangle _{a}$ and $\left\vert \alpha _{2}\right\rangle _{b}$ are
the coherent states for the vibrational modes along the x and y axes,
respectively, with $\alpha _{1}+\alpha _{2}=\sqrt{\varepsilon /\lambda }$.
The values of $\alpha _{1}$ and $\alpha _{2}$ depend on the initial state of
the vibrational modes. When they are initially in an eigenstate of the
parity operator $\Pi =(-1)^{(a^{\dagger }+b^{\dagger })(a+b)}$, the two
modes remain symmetric during the evolution and the two-mode state
components in the steady state are $\left\vert \alpha \right\rangle
_{a}\left\vert \alpha \right\rangle _{b}$ and $\left\vert -\alpha
\right\rangle _{a}\left\vert -\alpha \right\rangle _{b}$, with $\alpha =%
\frac{1}{2}\sqrt{\varepsilon /\lambda }$. We note the operator $\Pi $
commutes with the system Hamiltonian, so that the parity is conserved during
the process. When the parity is initially even, the vibrational modes will
finally evolves to the even two-mode cat state
\begin{equation}
\left\vert \psi _{+}\right\rangle ={\cal N}_{+}\left( \left\vert \alpha
\right\rangle _{a}\left\vert \alpha \right\rangle _{b}+\left\vert -\alpha
\right\rangle _{a}\left\vert -\alpha \right\rangle _{b}\right) ,
\end{equation}%
where ${\cal N}_{+}=\left( 2+2e^{-4\left\vert \alpha \right\vert
^{2}}\right) ^{-1/2}$. On the other hand, for the odd parity, the steady
state corresponds to the two-mode odd cat state
\begin{equation}
\left\vert \psi _{-}\right\rangle ={\cal N}_{-}\left( \left\vert \alpha
\right\rangle _{a}\left\vert \alpha \right\rangle _{b}-\left\vert -\alpha
\right\rangle _{a}\left\vert -\alpha \right\rangle _{b}\right) ,
\end{equation}%
where ${\cal N}_{-}=\left( 2-2e^{-4\left\vert \alpha \right\vert
^{2}}\right) ^{-1/2}$. We note that the sytem steady state remains unchanged
when the dephasing of the electronic degree of freedom is included in the
master equation. This is due to the fact that this degree of freedom is
finally in the ground state and not affected by dephasing.

\section{NUMERICAL SIMULATIONS}

To verify the validity of the proposed scheme, we perform a numerical
simulation of the fidelity of the system steady state with respect to the
target state. We first suppose that the target state is an even two-mode cat
state $\left\vert \psi _{+}\right\rangle $ with $\alpha =2$. To generate
such a state, we assume that the two vibrational modes are initially in the
vacuum state $\left\vert 0\right\rangle _{a}\left\vert 0\right\rangle _{b}$,
and the internal degree of freedom is initially in the ground state $%
\left\vert g\right\rangle $. With the coupling between the vibrational modes
and the reservoir being included, the master equation is%
\begin{eqnarray}
\frac{d\rho }{dt} &=&-i\left[ H_{i},\text{ }\rho \right] +\frac{\Gamma }{2}%
\left( 2S^{-}\rho S^{+}-S^{+}S^{-}\rho -\rho S^{+}S^{-}\right)  \nonumber \\
&&+\frac{\gamma _{a}}{2}\left( 2a\rho a^{\dagger }-a^{\dagger }a\rho -\rho
a^{\dagger }a\right)  \nonumber \\
&&+\frac{\gamma _{b}}{2}\left( 2b\rho b^{\dagger }-b^{\dagger }b\rho -\rho
b^{\dagger }b\right) ,
\end{eqnarray}%
where $\gamma _{a}$ and $\gamma _{b}$ are the decaying rates of the two
vibrational modes, respectively. We set $\eta _{x}=0.15$, $\eta _{y}=0.1$, $%
\varepsilon =16\lambda $, $\Gamma =100\lambda $, and $\gamma _{a}=\gamma
_{b}=0.0005\lambda $. To verify the validity of the Lamb-Dicke
approximation, here we retain the dominant higher-order terms ($j+l=1$) in
Eq~(3), so that the Hamiltonian is
\begin{eqnarray}
H_{i}&&=\left[ -\lambda \left( 1-\frac{\eta _{x}^{2}}{3}a^{\dagger }a\right)
a^{2}-\lambda \left( 1-\frac{\eta _{y}^{2}}{3}b^{\dagger }b\right)
b^{2}\right.  \nonumber \\&&
\left. -2\lambda \left( 1-\frac{\eta _{x}^{2}}{4}a^{\dagger }a-\frac{\eta
_{y}^{2}}{4}b^{\dagger }b\right) ab \right.  \nonumber \\&&
\left. + \varepsilon \left( 1-\frac{\eta _{x}^{2}
}{2}a^{\dagger }a-\frac{\eta _{y}^{2}}{2}b^{\dagger }b\right) \right]
e^{-i\phi _{0}}S^{+}+H.c.,
\end{eqnarray}

\begin{figure}
\centering
\includegraphics[width=3.5in]{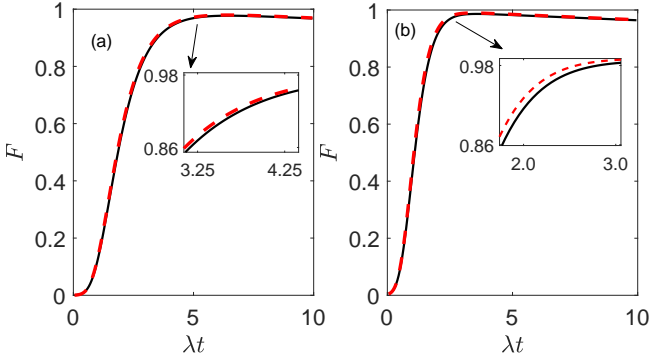}
\caption{(Color online) (a) Fidelity of the generated vibrational state to the
expected state as a function of $\lambda t$ with the initial state $%
\left\vert g\right\rangle \left\vert 0\right\rangle _{a}\left\vert
0\right\rangle _{b}$. The generated vibrational state is calculated by
numerically solving Eq. (14), with the parameters $\eta _{x}=0.15$, $\eta
_{y}=0.1$, $\varepsilon =16\lambda $, $\Gamma =100\lambda $, and $\gamma
_{a}=\gamma _{b}=0.0005\lambda $. The corresponding expected state is ${\cal N}%
_{+}(\left\vert \alpha \right\rangle _{a}\left\vert \alpha \right\rangle
_{b}+\left\vert -\alpha \right\rangle _{a}\left\vert -\alpha \right\rangle
_{b})$ with $\alpha =2$. The black-solid and red-dashed lines represent the results
with and without consideration of the dominant higher-order terms of the
Hamiltonian, respectively. (b) Fidelity of the generated vibrational state
to the expected state as a function of $\lambda t$ with the initial state $
\left\vert g\right\rangle \left( \left\vert 1\right\rangle _{a}\left\vert
0\right\rangle _{b}+\left\vert 0\right\rangle _{a}\left\vert 1\right\rangle
_{b}\right) /\sqrt{2}$. The corresponding expected state is ${\cal N}%
_{-}(\left\vert \alpha \right\rangle _{a}\left\vert \alpha \right\rangle
_{b}-\left\vert -\alpha \right\rangle _{a}\left\vert -\alpha \right\rangle
_{b})$ with $\alpha =2$. The unit of this figure is $\lambda$.  }
\end{figure}
Fig. 1(a) shows the fidelity (black-solid line), defined as $F=\left\langle \psi
_{+}\right\vert \rho _{a,b}\left\vert \psi _{+}\right\rangle $, as a
function of the evolution time, where $\rho _{a,b}$ is the reduced density
operator of the two vibrational modes, obtained by tracing the total system
density operator over the internal degree of freedom. The numerical result
shows that the system approaches the steady state when $\lambda t\simeq 6.5$;
at this time the fidelity is $F\simeq 0.977$. In Fig. 1(b), we display numerical
simulation of the evolution of the fidelity for the obtained vibrational
state to the ideal odd cat state $\left\vert \psi _{-}\right\rangle $ with
the initial state $\left\vert g\right\rangle \left( \left\vert
1\right\rangle _{a}\left\vert 0\right\rangle _{b}+\left\vert 0\right\rangle
_{a}\left\vert 1\right\rangle _{b}\right) /\sqrt{2}$, where $\left\vert
1\right\rangle _{a}$ and $\left\vert 1\right\rangle _{b}$ denote the
one-phonon Fock state for modes a and b, respectively. In this case the
corresponding fidelity is about $F\simeq0.986$ at the time $\lambda t\simeq 3.5$. Compared
to the cases without taking the higher-order terms into consideration
(red-dashed lines), the steady state fidelities are decreased by only about $0.25\%$,
confirming the validity of the approximation.

To further demonstrate the production of the desired entanglement in each
steady state, we calculate the corresponding joint Wigner function of the two bosonic
modes, defined as the joint quasiprobability
distribution for these modes in four-dimensional phase space
\begin{equation}
W(\beta ,\chi )=\frac{4}{\pi ^{2}}\left\langle D_{a}(\beta )(-1)^{a^{\dagger
}a}D_{a}^{\dagger }(\beta )D_{b}(\chi )(-1)^{b^{\dagger }b}D_{b}^{\dagger
}(\chi )\right\rangle ,
\end{equation}%
where $D_{a}(\beta )=e^{\beta a^{\dagger }-\beta ^{\ast }a}$ and $%
D_{b}(\chi )=e^{\chi b^{\dagger }-\chi ^{\ast }b}$ are displacement
operators for the vibrational modes along the x and y axes, respectively,
with $\beta $ and $\chi $ being complex parameters, which define the
coordinates in the joint phase space [24]. The dominant higher-order terms
in the Hamiltonian are included in the calculation of $W(\beta ,\chi )$. The
plane-cut of the Wigner function along the Im($\beta $)-Im($\chi $) axes for
the ideal even two-mode cat state with $\alpha =2$ is shown in Fig. 2 (a),
while that for the steady state obtained with the initial state $\left\vert g\right\rangle \left\vert 0\right\rangle
_{a}\left\vert 0\right\rangle _{b}$ and the above mentioned parameters at the time $\lambda t=7$
 is shown in Fig. 2 (b). The fringes with
alternating positive and negative values on the Im($\beta $)-Im($\chi $)
plane-cut are the signatures of quantum interference between the two
quasiclassical components. The plane-cuts of the Wigner functions along the
Im($\beta $)-Im($\chi $) axes for the ideal odd two-mode cat state with $
\alpha =2$ and for the steady state obtained with the initial state $\left\vert
g\right\rangle \left( \left\vert 1\right\rangle _{a}\left\vert
0\right\rangle _{b}+\left\vert 0\right\rangle _{a}\left\vert 1\right\rangle
_{b}\right) /\sqrt{2}$ at the time $ \lambda t=4$ are shown in Fig. 3 (a) and (b), respectively. As
expected, for each initial state the plane-cuts of the Wigner function along
the Im($\beta $)-Im($\chi $) axes obtained from the steady state is in well
agreement with the result for the ideal cat state, and the interference
features associated with the two initial states are complementary.

\begin{figure}
\centering
\includegraphics[width=3.5in]{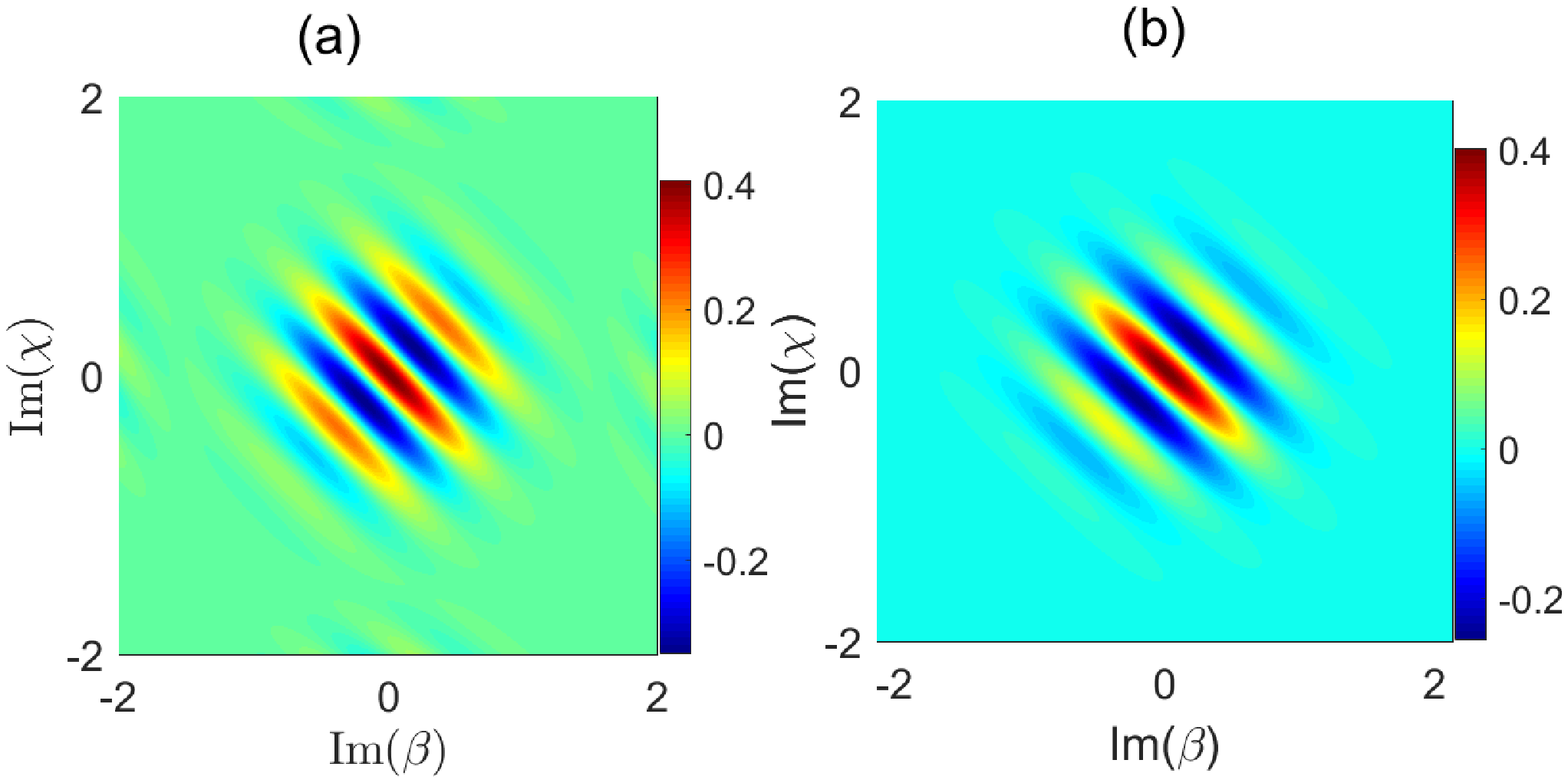}
\caption{(Color online)  (a) Plane-cut of the two-mode Wigner function $W(\beta
,\chi )$ along the Im($\beta $)-Im($\chi $) axes for the ideal cat state $%
{\cal N}_{+}(\left\vert \alpha \right\rangle _{a}\left\vert \alpha
\right\rangle _{b}+\left\vert -\alpha \right\rangle _{a}\left\vert -\alpha
\right\rangle _{b})$ with $\alpha =2$. (b) Plane-cut of $W(\beta ,\chi )$
along the Im($\beta $)-Im($\chi $) axes obtained by numerically solving Eq.
(8) with the initial state $\left\vert g\right\rangle \left\vert
0\right\rangle _{a}\left\vert 0\right\rangle _{b}$. The parameters are the
same as those in Fig. 1, and the dominant higher-order terms in the
Hamiltonian are included. The unit of this figure is $\lambda$. }
\end{figure}

\begin{figure}
\centering
\includegraphics[width=3.5in]{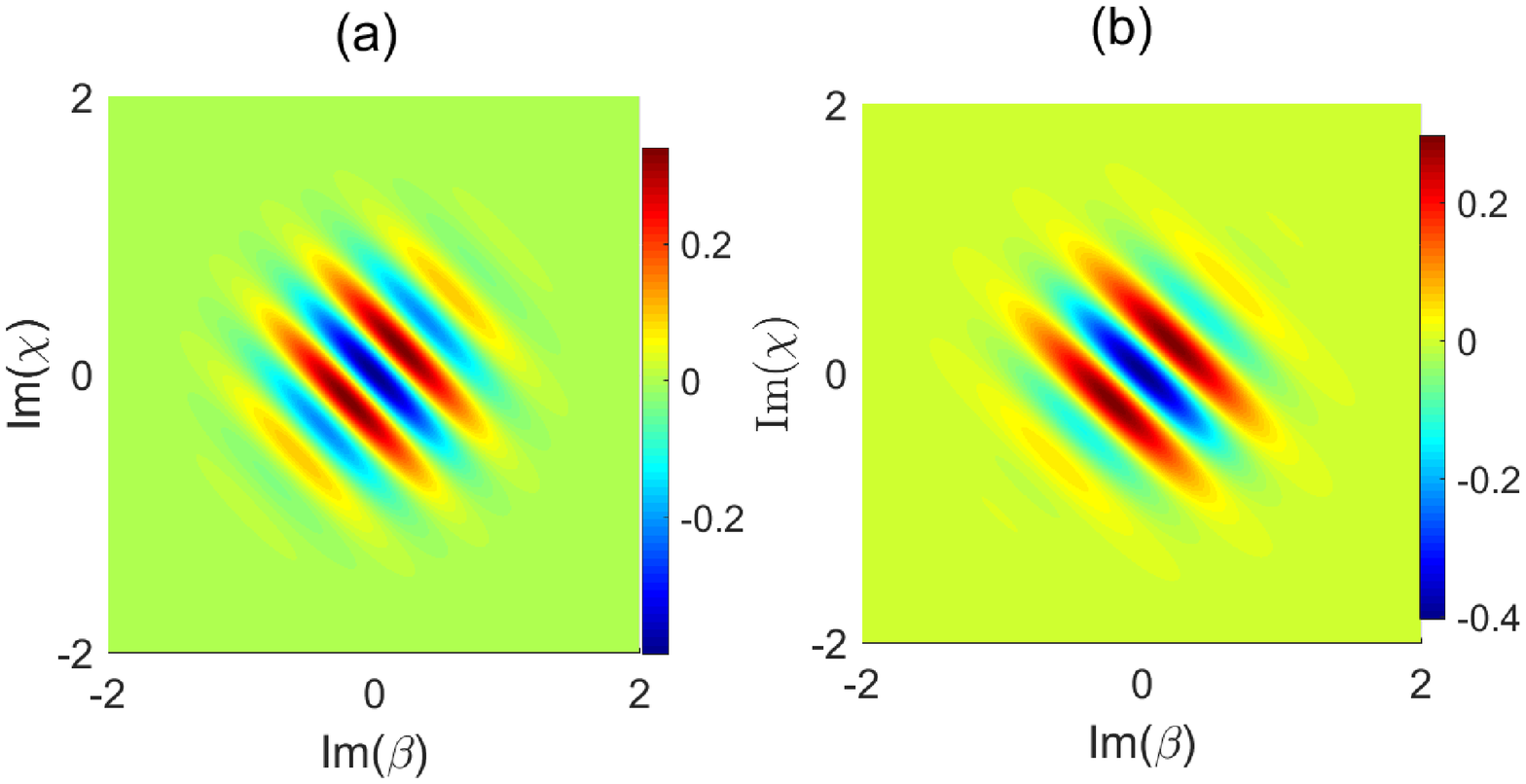}
\caption{(Color online) (a) Plane-cut of $W(\beta ,\chi )$ along the Im($\beta
$)-Im($\chi $) axes for the ideal cat state ${\cal N}_{-}(\left\vert \alpha
\right\rangle _{a}\left\vert \alpha \right\rangle _{b}-\left\vert -\alpha
\right\rangle _{a}\left\vert -\alpha \right\rangle _{b})$ with $\alpha =2$.
(b) Plane-cut of $W(\beta ,\chi )$ along the Im($\beta $)-Im($\chi $) axes
for the vibrational modes obtained by simulating the master equation with
the initial state $\left\vert g\right\rangle \left( \left\vert
1\right\rangle _{a}\left\vert 0\right\rangle _{b}+\left\vert 0\right\rangle
_{a}\left\vert 1\right\rangle _{b}\right) /\sqrt{2}$. The parameters are the
same as those in Fig. 1, and the dominant higher-order terms in the
Hamiltonian are included. The unit of this figure is $\lambda$. }
\end{figure}

\section{GENERATION OF THREE-MODE CAT STATES}

We now turn to the case that the ion is in a three-dimensional anisotropic
trap, with the vibrational frequencies along the x, y, and z axes bieng $%
\omega _{x}$, $\omega _{y}$, and $\omega _{z}$, respectively. The ion is
driven by six laser beams of frequencies $\omega _{0}-2\omega _{x}$, $\omega
_{0}-2\omega _{y}$, $\omega _{0}-2\omega _{z}$, $\omega _{0}-\omega
_{x}-\omega _{y}$, $\omega _{0}-\omega _{y}-\omega _{z}$, $\omega
_{0}-\omega _{x}-\omega _{z}$, and $\omega _{0}$. The first three laser
beams are aligned along the x and y axes, the fourth is on the x-y plane and
at angle of $\pi /4$ to x axis, the fifth is on the y-z plane and at an
angle of $\pi /4$ to y axis, the sixth is on the x-z plane and at angle of $%
\pi /4$ to x axis, and the last one is on the x-y plane and at an angle of $%
-\pi /4$ to x axis. In the rotating-wave approximation, the Hamiltonian for
this system is given by
\begin{eqnarray}
&&E^{+}(x,t) =E_1e^{-i[(\omega _0-2\omega _x)t-k_1x]}+E_2e^{-i[(\omega
_0-2\omega _y)t-k_2y]}  \nonumber \\&&
+E_3e^{-i[(\omega _0-2\omega _y)t-k_3z]}+E_4e^{-i[(\omega _0-\omega_x-\omega _y)t-k_4(x+y)/\sqrt{2}]}  \nonumber \\&&
+E_5e^{-i[(\omega _0-\omega _y-\omega _z)t-k_5(y+z)/\sqrt{2}]}\nonumber\\&&
+E_6e^{-i[(\omega _0-\omega _x-\omega _z)t-k_6(x+z)/\sqrt{2}]}  \nonumber\\&&
+E_0e^{-i[\omega _0t-k_0(x-y)/\sqrt{2}]}.
\end{eqnarray}
We here have assumed that the phase for each laser beam is 0. In the
resolved sideband limit and in the Lamb-Dicke regime, the effective
Hamiltonian for the system is
\begin{eqnarray}
H_i&&=\left[ \frac{-\eta _x^2}2e^{-\eta _x^2/2}\Omega _1a^2-\frac{\eta _y^2}2
e^{-\eta _y^2/2}\Omega _2b^2\right.  \nonumber \\&&
-\frac{\eta _z^2}2e^{-\eta _z^2/2}\Omega _3c^2-\frac{\eta _x\eta _y}2
e^{-(\eta _x^2+\eta _y^2)/4}\Omega _4ab \nonumber \\&&
-\frac{\eta _y\eta _z}2e^{-(\eta _y^2+\eta _z^2)/4}\Omega _5bc-\frac{\eta
_x\eta _z}2e^{-(\eta _x^2+\eta _z^2)/4}\Omega _6ac \nonumber \\&&
\left. +e^{-(\eta _x^2+\eta _y^2)/4}\Omega _0\right] S^{+}+H.c.,
\end{eqnarray}
where $c$ is the annihilation operator for the vibrational mode along the z
axis. We choose the amplitudes of the lasers appropriately so that
\begin{eqnarray}
\eta _x^2e^{-\eta _x^2/2}\Omega _1 =\eta _y^2e^{-\eta _y^2/2}\Omega
_2=\eta _z^2e^{-\eta _z^2/2}\Omega _3=\lambda ,  \nonumber \\
\eta _x\eta _ye^{-(\eta _x^2+\eta _y^2)/4}\Omega _4 =\eta _y\eta
_ze^{-(\eta _y^2+\eta _z^2)/4}\Omega _5 \nonumber \\=\eta _x\eta _ze^{-(\eta
_x^2+\eta _z^2)/4}\Omega _6=2\lambda ,\nonumber \\
\end{eqnarray}
then we obtain

\begin{equation}
H_i=[-\lambda (a+b+c)^2+\varepsilon ]S^{+}+H.c.
\end{equation}

When the decaying of the electronic state is included, the system dynamics
is governed by the master equation of form of Eq. (8). The system steady
state is again of the form of Eq. (10), where $\left| \psi \right\rangle $
satisfies
\begin{equation}
\lambda (a+b+c)^2\left| \psi \right\rangle =\varepsilon \left| \psi
\right\rangle .
\end{equation}
When the vibrational modes are initially in an eigenstate of the parity
operator $\Pi =(-1)^{(a^{\dagger }+b^{\dagger }+c^{\dagger })(a+b+c)}$, $%
\left| \psi \right\rangle $ is a superposition of $\left| \alpha
\right\rangle _a\left| \alpha \right\rangle _b\left| \alpha \right\rangle _c$
and $\left| -\alpha \right\rangle _a\left| -\alpha \right\rangle _b\left|
-\alpha \right\rangle _c$, with $\alpha =\frac 13\sqrt{\varepsilon /\lambda }
$. The relative superposition coefficient of the two components in the
steady state depends on the initial parity. When these modes are initially
in the vacuum state $\left| 0\right\rangle _a\left| 0\right\rangle _b\left|
0\right\rangle _c$, they will finally evolves to the even three-mode cat
state
\begin{equation}
\left| \psi _{+}\right\rangle ={\cal N}_{+}\left( \left| \alpha
\right\rangle _a\left| \alpha \right\rangle _b\left| \alpha \right\rangle
_c+\left| -\alpha \right\rangle _a\left| -\alpha \right\rangle _b\left|
-\alpha \right\rangle _c\right) ,
\end{equation}
where ${\cal N}_{+}=\left( 2+2e^{-8\left| \alpha \right| ^2}\right) ^{-1/2}$%
. On the other hand, if the vibrational modes are initially in the state $%
\left( \left| 1\right\rangle _a\left| 0\right\rangle _b\left| 0\right\rangle
_c+\left| 0\right\rangle _a\left| 1\right\rangle _b\left| 0\right\rangle
_c+\left| 0\right\rangle _a\left| 0\right\rangle _b\left| 1\right\rangle
_c\right) /\sqrt{3}$, the steady state corresponds to the three-mode odd cat
state
\begin{equation}
\left| \psi _{-}\right\rangle ={\cal N}_{-}\left( \left| \alpha
\right\rangle _a\left| \alpha \right\rangle _b\left| \alpha \right\rangle
_c-\left| -\alpha \right\rangle _a\left| -\alpha \right\rangle _b\left|
-\alpha \right\rangle _c\right) ,
\end{equation}
where ${\cal N}_{-}=\left( 2-2e^{-8\left| \alpha \right| ^2}\right) ^{-1/2}$.

\section{SUMMARY}

In conclusion, we have proposed a scheme for producing entangled coherent
states of the vibrational modes of an ion trapped in a two-dimensional
anisotropic harmonic well. In our scheme, the ion is driven by four laser
beams, one of which is used to directly coupling the two electronic states,
while the others serve for coupling the electronic transition with the
vibrational modes. In the Lamb-Dicke limit and with appropriate setting of
the laser parameters, the combination of the unitary dynamics and
spontaneous emission will evolve the system to a steady state, given by the
product of the electronic ground state with a two-mode vibrational state.
When the two vibrational modes are initially in a Fock state, their steady
state corresponds to a cat state. Numerical simulations confirm the validity
of the proposed scheme. We further show that the method can be generalized
to realization of three-mode cat states for an ion trapped in a
two-dimensional anisotropic harmonic well. We note the idea may also be
applied to generation and stabilization of entangled coherent states for two
cavity modes by coupling them to an atom or an artificial atom.

This work was supported by the National Natural Science Foundation of China
under Grant No. 11674060 and No. 11705030, the Natural Science Foundation of Fujian Province
under Grant No. 2016J01018, and the Fujian Provincial Department of
Education under Grant No. JZ160422.


\begin{references}
\bibitem{} E. Schr\"odinger, Naturwissenschaften 23, 807 (1935).

\bibitem{} S. Del\'eglise, I. Dotsenko, C. Sayrin, J. Bernu, M. Brune, J.-M.
Raimond, and S. Haroche, Nature 455, 510 (2008).

\bibitem{} Z. Leghtas et al., Phys. Rev. Lett. 111, 120501 (2013).

\bibitem{} M. Mirrahimi et al., New J. Phys. 16, 045014 (2014).

\bibitem{} N. Ofek et al., Nature (London) 536, 441 (2016).

\bibitem{} V. V. Albert et al., Phys. Rev. Lett. 116, 140502 (2016).

\bibitem{} M. Brune et al., Phys. Rev. Lett. 77, 4887 (1996).

\bibitem{} M. Hofheinz et al., Nature (London) 459, 546 (2009).

\bibitem{} S.-B. Zheng et al., Phys. Rev. Lett. 115, 260403 (2015).

\bibitem{} B. Vlastakis, G. Kirchmair, Z. Leghtas, S. E. Nigg, L. Frunzio,
S. M. Girvin, M. Mirrahimi, M. H. Devoret, and R. J. Schoelkopf, Science
342, 607 (2013).

\bibitem{} L. Sun et al., Nature 511, 444C448 (2014).

\bibitem{} B. Vlastakis et al., Nat. Commun. 6, 8970 (2015).

\bibitem{} K. Liu et al., Sci. Adv. 3, e1603159 (2017).

\bibitem{} C. Monro, D. M. Meekhof, B. E. King and D. J. Wineland, Science
272, 1131 (1996).

\bibitem{} C. Hempel et al., Nat. Photonics 7, 630 (2013).

\bibitem{} K. G. Johnson, J. D. Wong-Campos, B. Neyenhuis, J. Mizrahi, and
C. Monroe, Nature Communications 8, 697 (2017).

\bibitem{} B. C. Sanders, Phys. Rev. A 45, 6811 (1992).

\bibitem{} C. Chai, Phys. Rev. A 46, 7187 (1992).

\bibitem{} B. C. Sanders, J. Phys. A: Math. Theor. 45, 244002 (2012).

\bibitem{} J. Joo, W. J. Munro, and T. P. Spiller, Phys. Rev. Lett. 107,
083601 (2011).

\bibitem{} A. Sarlette, Z. Leghtas, M. Brune, J. M. Raimond, and P. Rouchon,
Phys. Rev. A 86, 012114 (2012).

\bibitem{} C. Arenz, C. Cormick, D. Vitali, and G. Morigi, J. Phys. B: At.
Mol. Opt. Phys. 46, 224001 (2013).

\bibitem{} M. Mamaev, L. C. G. Govia, and A. A. Clerk, arXiv:1711.06662.

\bibitem{} C. Wang et al., Science 352, 1087 (2016).

\bibitem{} D. M. Meekhof, C. Monroe, B. E. King, W. M. Itano, and D.J.
Wineland, Phys. Rev. Lett. 76, 1796 (1996).

\bibitem{} B. DeMarco, A. Ben-Kish, D. Leibfried, V. Meyer, M. Rowe, B.M.
Jelenkovi\'{c}, W.M. Itano, J. Britton, C. Langer, T. Rosenband, and D.J.
Wineland, Phys. Rev. Lett. 90, 037902 (2003).

\bibitem{} J. D. Jost, J. P. Home, J. M. Amini, D. Hanneke, R. Ozeri, C.
Langer, J. J. Bollinger, D. Leibfried, and D. J. Wineland, Nature 459, 683
(2009).

\bibitem{} S. -C. Gou, J. Steinbach and P. L. Knight, Phys. Rev. A 54, R1014
(1996).

\bibitem{} S. -C. Gou, J. Steinbach and P. L. Knight, Phys. Rev. A 54, 4315
(1996).

\bibitem{} C. C. Gerry, S. -C. Gou and J. Steinbach, Phys. Rev. A55, 630
(1997).

\bibitem{} S. -B. Zheng and G. -C. Guo, Quantum Semiclass. Opt. 10, 441
(1998).

\bibitem{} S. -B. Zheng, Phys. Rev. A 58, 761 (1998).

\bibitem{} C. C. Gerry, Phys. Rev. A55, 2478 (1997).

\bibitem{} S. A. Gardiner, J. I. Cirac, and P. Zoller, Phys. Rev. A 55, 1683
(1997).

\bibitem{} B. Kneer and C. K. Law, Phys. Rev. A 57, 2096 (1998).

\bibitem{} G. Drobny, B. Hladky, and V. Buzek, Phys. Rev. A 58, 2481 (1998).

\bibitem{} S.-B. Zheng, Phys. Rev. A. 63, 015801 (2001).

\bibitem{} X.-B Zou, K Pahlke, and W Mathis, Phys. Rev. A 65, 045801 (2002).

\bibitem{} W. Vogel and R. L. de Matos Filho, Phys. Rev. A 52, 4214 (1995).

\bibitem{} R. L. de Matos Filho and W. Vogel, Phys. Rev. Lett. 76, 608
(1996).

\bibitem{} N. B. An and T. M. Duc, Phys. Rev. A 66, 065401 (2002).
\end{references}
\end{document}